\documentclass[12pt]{iopart}
\usepackage{graphicx}
\usepackage{epstopdf}
\usepackage{bbm}

\newcommand{\Lg}{L_{\rm geo}}
\newcommand{\Lj}{L_{\rm j}}
\newcommand{\Lt}{L_{\rm tot}}

\newcommand{\Peo}{\Phi_{\rm e0}}
\newcommand{\mur}{\mu_{\rm r,eff}}

\begin{document}

\title{A one-dimensional tunable magnetic metamaterial}

\author{S Butz$^1$, P Jung$^1$, L V Filippenko$^{2,3}$, V P Koshelets$^{2,3}$ and A~ V~ Ustinov$^{1,3}$}

\address{$^1$ Physikalisches Institut, Karlsruhe Institute of Technology,
D-76131 Karlsruhe, Germany}

\address{$^2$ Kotel'nikov Institute of Radio Engineering and Electronics (IREE RAS), Moscow 125009, Russia}

\address{$^3$ National University of Science and Technology MISIS, Moscow 119049, Russia}

\ead{s.butz@kit.edu}

\begin{abstract}
We present experimental data on a one-dimensional superconducting meta\-ma\-terial that is tunable over a broad frequency band. The basic building block of this magnetic thin-film medium is a single-junction (rf-) superconducting quantum interference device (SQUID). Due to the nonlinear inductance of such an element, its resonance frequency is tunable in situ by applying a dc magnetic field. We demonstrate that this  results in tunable effective parameters of our metamaterial consisting of 54 SQUIDs. In order to obtain the effective magnetic permeability $\mur$ from the measured data, we employ a technique that uses only the complex transmission coefficient $S_{21}$.

\end{abstract}

\maketitle

\section{Introduction}

Losses and the strong limitation to a narrow frequency band are the main challenges when designing metamaterials that are made of conventional resonant structures. It has been shown  that losses in metamaterials, working in and below the THz frequency range, can be greatly reduced when metallic structures are replaced by superconducting ones \cite{ricci2005}. Additionally, superconducting meta-atoms exhibit an intrinsic tunability of their resonance frequency by magnetic field and temperature \cite{ricci2007, gu2010}. However, in both cases, the tunability arises from a suppression of the superconducting order parameter, i.e. the density of Cooper pairs. Thus, by tuning the resonance frequency, the quality factor of the resonance is changed as well.

In this work, we demostrate a one-dimensional metamaterial that employs superconducting quantum interference devices (SQUIDs) as meta-atoms, based on a theoretical idea introduced and further investigated in \cite{lazarides2007, du2006} and \cite{gabitov2012}. The SQUID can be considered as a split ring resonator (SRR) \cite{smith2000} that includes a Josephson junction. The tunability of its resonance frequency arises from the inductance of the Josephson junction that is tunable by a very weak dc magnetic field and does not come at the cost of suppression of superconductivity. The magnetic field tunability of the resonance frequency of such a SQUID meta-atom has been experimentally verified in \cite{jung13}. This reference serves as basis for the microwave properties of a single SQUID.

\section{The rf-SQUID}\label{sec:rfsquid}

\begin{figure}[htpb]
\centering\includegraphics[width = 1\textwidth]{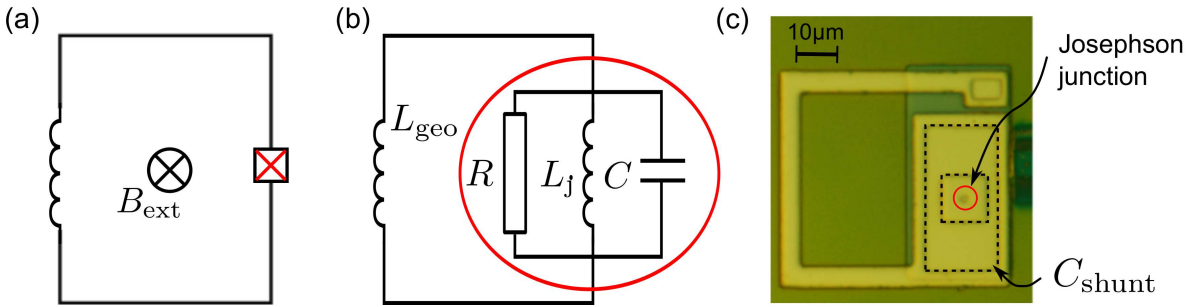}
\caption{(a) Sketch of an rf-SQUID. The red cross symbolizes the Josephson junction. (b) Electric equivalent circuit for an rf-SQUID in the small signal approximation. $\Lg$ is the geometric inductance of the SQUID loop. The red circle indicates the electric circuit model for the junction. $R$ represents the resistance due to a quasiparticle current, $\Lj$ is the Josphehson junction and $C$ stands for the capacitance between the superconducting electrodes. (c) Optical micrograph of the rf-SQUID.}
\label{fig:squid_waveguide}
\end{figure}

The basic building block of the metamaterial used in this work is a single junction (rf-)SQUID. Such an rf-SQUID consists of a superconducting loop interrupted by a Josephson junction (cf. figure\,\ref{fig:squid_waveguide}(a)). The junction is indicated by a red cross.

\noindent
Within a small signal approximation, the Josephson junction can be considered as a nonlinear inductor \cite{likharev, tinkham}. When placed inside a superconducting loop, this so-called Josephson inductance $\Lj$ is tunable by a magnetic field. In addition to $\Lj$, the geometric inductance of the loop $\Lg$ contributes to the total inductance $\Lt$ of the rf-SQUID. The full equivalent electric circuit is depicted in figure\,\ref{fig:squid_waveguide}(b). The red circle marks the electric circuit analogue of the Josephson junction for which the resistively capacitively shunted junction model is used \cite{tinkham}. Like the SRR, the rf-SQUID can be interpreted as an LC-oscillator. Unlike the SRR however, the total inductance and thus the resonance frequency of the rf-SQUID is tunable, assuming the ac magnetic field component is small. 

Figure\,\ref{fig:squid_waveguide}(c) shows an optical micrograph of the single SQUID. The SQUID and its junction are fabricated using a $\rm Nb$/$\rm AlO_x$/$\rm Nb$ trilayer process. The Josephson junction is circular with a diameter of $1.6\,\mu$m, its critical current $I_c = 1.8\,\mu$A. From this value, the zero field Josephson inductance is calculated to be $\Lj = 183$\,pH. This value is approximately twice as large as the geometric inductance of the loop $\Lg = 82.5$\,pH. Thus, the rf-SQUID considered in this work is nonhysteretic \cite{likharev, tinkham}. The junction is shunted with an additional parallel plate capacitor with a capacitance $C_{\rm shunt} = 2.0$\,pF which is two orders of magnitude larger than the intrinsic capacitance of the Josephson junction. Due to this shunt capacitor, the resonance frequency of the rf-SQUID is reduced and tunable between approximately 9\,GHz and 15\,GHz.

\section{The SQUID Metamaterial}
\begin{figure}[htbp]
 \centering
 \includegraphics[width = 1\textwidth]{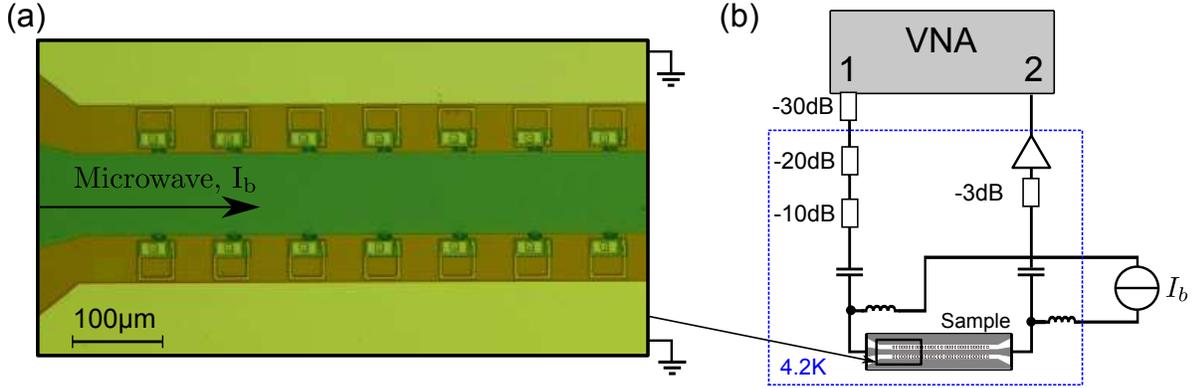}
 \caption{(a) Optical micrograph of part of the CPW containing a chain of rf-SQUIDs in each gap. (b) Measurement setup including the vector network analyzer (VNA), the biasTees, attentuation and cryogenic amplifier.}
 \label{fig:setup}
\end{figure}

Two one-dimensional arrays of 27 rf-SQUIDs are placed inside the gaps of a coplanar waveguide (CPW) as shown in figure\,\ref{fig:setup}(a). The electromagnetic wave travels along the waveguide. In addition to the microwave signal, a dc current is applied along the central conductor, creating the dc magnetic field used for tuning the resonance frequency. Due to the waveguide geometry, the SQUIDs are oriented with their loop area perpendicular to the magnetic field. 

The pitch between neighboring SQUIDs in this sample is $92\,\mu$m, which is twice the width of the single SQUID and much larger than the distance between each SQUID and the central conductor of the waveguide. Therefore, the inductive coupling between adjacent SQUIDs is approximately one order of magnitude smaller than the coupling to the CPW and can be neglected. 

\noindent The CPW is connected to a vector network analyzer (VNA). The full experimental setup is depicted in figure\,\ref{fig:setup}(b). Rigorous magnetic shielding (not shown in the picture) proved to be crucial to protect the sample from stray magnetic fields. The sample inside the cryoperm magnetic shield, part of the attenuation and the amplifier are placed at liquid helium temperature, $T = 4.2$\,K.

\section{Experimental Results}
\begin{figure}[htbp]
\centering
 \includegraphics[width = .55\textwidth]{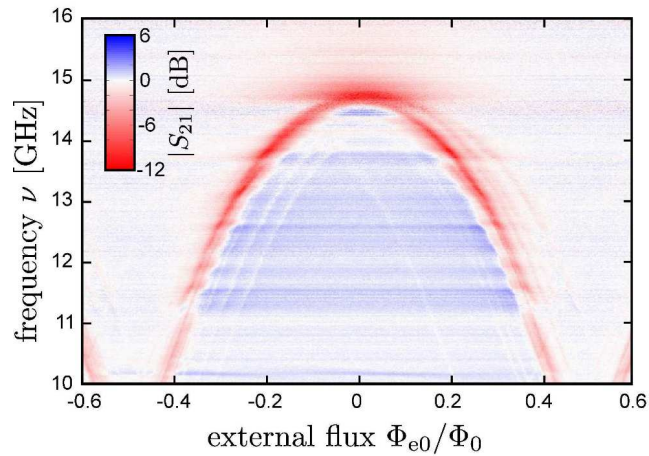}
 \caption{Measured transmission magnitude $\left |S_21\right |$ depending on frequency and magnetic flux $\Peo/\Phi_0$. }
 \label{fig:trans_magn}
\end{figure}

We measure the complex transmission through the CPW ($S_{21}$) as a function of frequency~ $\nu$ and magnetic flux~ $\Peo$. The microwave power at the sample is ap\-prox\-i\-mate\-ly $P\approx -90$\,dBm, including losses in the coaxial cables. The calibration of the measurement is done by applying a flux of $\Peo = \Phi_{\rm cal}=\Phi_0/2$. At this flux value the resonance frequency is shifted to its lowest possible value, which lies between 9 and 10\,GHz. The built-in ``thru'' calibration function of the VNA is used to substract the corresponding reference data from the rest of the measurement (see also supplimentary material). The resulting transmission magnitude for such a measurement is presented in figure\,\ref{fig:trans_magn}. For clarity, only data above 10\,GHz are shown.  

Note that for the presentation of the data in this section, the more natural quantity of magnetic flux per SQUID $\Peo$ is used instead of magnetic field. For our SQUIDs, a flux of one flux quantum $\Phi_0 = h/(2e) = 2.07\times
 10^{-15}$\,Vs corresponds to a field of $1.2\,\mu$T.

\noindent The resonance dip and its tunability with magnetic flux is clearly visible (low transmission is shown in red). It is the same behaviour that was already observed in \cite{jung13} for a single rf-SQUID. Here, however, the resonance is much more pronounced due to the increased number of SQUIDs. It should also be noted, that the shape and, more importantly, the depth of the resonance do not change considerably over a wide frequency band between 13\,GHz and 14.5\,GHz. The distortions in the onset and shape of the resonance dip, for example at 13.6\,GHz and at 14.5\,GHz, are due to parasitic sample holder resonances that couple to the SQUIDs. They are also faintly visible as flux independent stripes far away from resonance.

The few separate lines, that are shifted in the flux axis with respect to the main resonance line, are lines from a few indivual SQUIDs. Their magnetic flux bias is different from the flux bias of the main part of the SQUID array. This is most probably due to trapped Abrikosov vortices either in the central conductor or in the superconducting material of the SQUIDs themselves. 

\begin{figure}[htbp]
\centering\includegraphics[width = 1\textwidth]{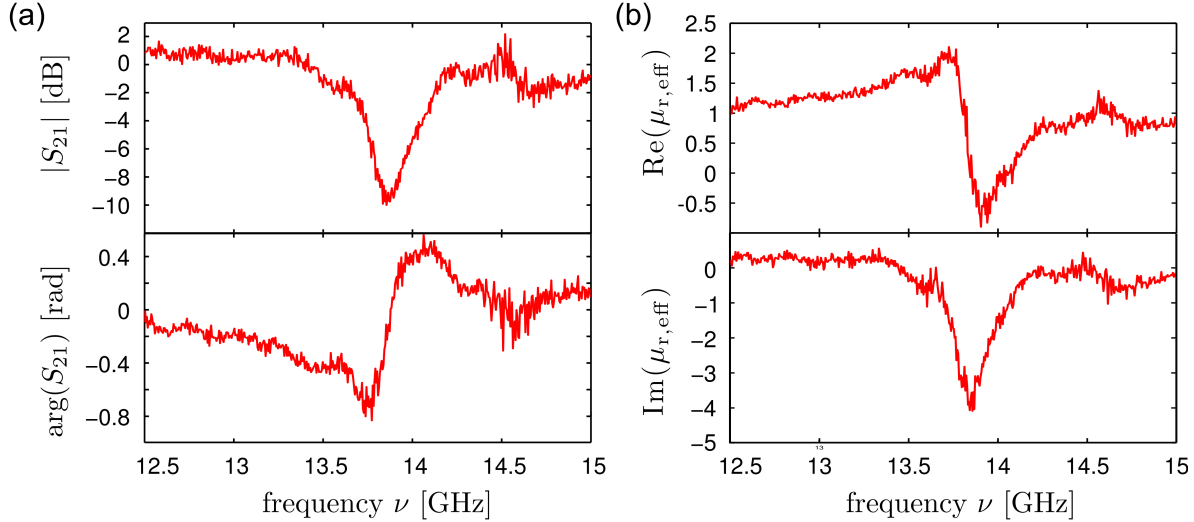} 
\caption{(a) Transmission magnitude (top) and phase (bottom) at an external flux of $\Peo = -0.185\Phi_0$. (b) Real (top) and imaginary (bottom) part of the effective magnetic permeability $\mur$ calculated from the transmission data shown in (a). It should be noted that we employ $\mur = \rm Re(\mur) + \rm i\, Im(\mur)$, opposite to the commonly used definition $\mu = \mu' - i\,\mu''$.}
\label{fig:ST}
\end{figure}

\noindent Figure\,\ref{fig:ST}(a) shows a zoom to a smaller frequency range for a fixed flux value of external flux $\Peo = -0.185\Phi_0$. It allows a more detailed look at transmission magnitude and phase around the resonance frequency of $\nu_0 = 13.88$\,GHz. We observe that the resonance dip is indeed one collective resonance dip, no single SQUID lines are distiguishable. The steep rise of the phase dependence underlines the collective behaviour. However, as expected from the weak coupling between adjacent SQUIDs, there is no mutual synchronization. The quality factor of this resonance is $Q_{\rm collective} = 100$. When we compare this value to the quality factor of a single SQUID line (we used one of the weak stray lines mentioned earlier), the quality factor of the single SQUID is more than twice as large $Q_{\rm single} = 215$. A detailed investigation of the quality factors, however, is hindered by the very shallow resonance of the single SQUID which is less than 1dB.

By using a method described in the next section, we are able retrieve the effective magnetic permeability $\mur$ from the measured transmission data depicted in figure\,\ref{fig:trans_magn}. In figure\,\ref{fig:ST}(b), real and imaginary part of the effective magnetic permeability $\mur$ are plotted for the same frequency interval and flux value as the transmission data in Fig.\ref{fig:ST}(a). The magnetic permeability shows a frequency dependence typical for a metamaterial consisting of resonant elements \cite{pendry1999}. 
Upon approaching the resonance frequency, the real part of $\mur$ increases from unity to almost 2. When reaching $\nu_0$ it drops drastically to values below zero, which is due to the change of phase between incoming signal and magnetic response of the SQUIDs. From there it increases again to unity. The comparatively slow increase of Re($\mur$) between 12.5\,GHz and 13.5\,GHz as well as the small dip slightly above 13.5\,GHz is due to a sample holder resonance at 13.6\,GHz which couples to the SQUIDs and affects the resonance. The deviation from zero of the imaginary part of $\mur$ reflects the increased losses at resonance. 

\begin{figure}[htpb]
 \centering \includegraphics[width = 1\textwidth]{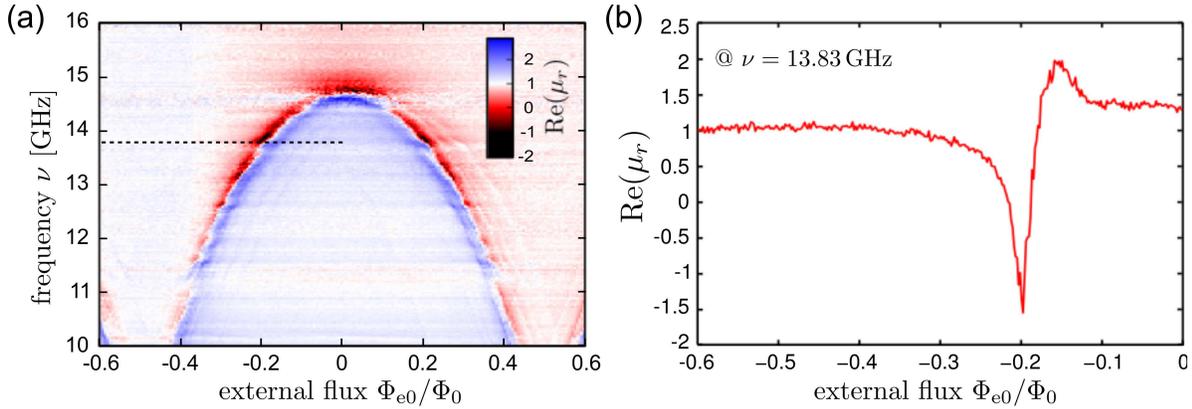}
 \caption{(a) Frequency and flux dependent real part of the effective magnetic permeability Re($\mur$) calculated from the transmission data shown in figure\,\ref{fig:trans_magn}. The black dashed line indicates the cut shown in (b). (b) Flux dependence of Re($\mur$) at a frequency $\nu = 13.83$. Note that only negative flux values are shown.}
 \label{fig:mur2d}
\end{figure}

\noindent The full frequency and flux dependence of the real part of the effective magnetic permeability is shown in colour scale in figure\,\ref{fig:mur2d}(a). The frequency dependent behaviour of Re($\mur$) described above, extents qualitatively over the full frequency and flux range of the resonance curve displayed in figure\,\ref{fig:trans_magn}. 
The data show clearly how the SQUID metamaterial enables us to reach any desired value of $\mur$ for a given frequency. The minimum and maximum achieved values are Re$(\mur)_{\rm min} = -2$ and Re$(\mur)_{\rm max} = 3$, respectively. Unfortunately, due to sample holder resonances, especially around 13.6\,GHz and close to the maximum of the curve at approximately 14.5\,GHz, these values cannot be reached for every measured frequency. Figure\,\ref{fig:mur2d}(b) shows the real part of the magnetic permeability for a fixed frequency of $\nu = 13.83$\,GHz, highlighted by the black dashed line in figure\,\ref{fig:mur2d}(a). For this frequency, Re$(\mur)$ can be tuned to any value between -1.5 and +2.

\section{Effective Parameter Retrieval}\label{sec:retrieval}
\begin{figure}[thbp]
 \centering \includegraphics[width = \textwidth]{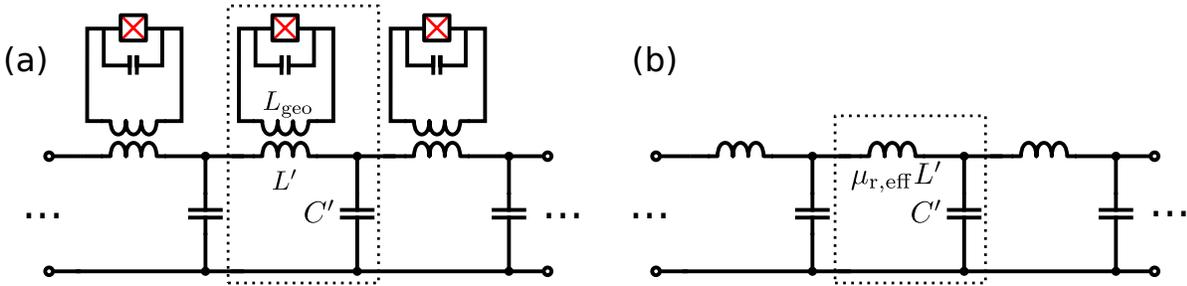}
 \caption{Electrical model of the circuit for one SQUID per unit cell. The unit cells in both pictures are indicated by a dashed box. (a) The CPW is modeled as a transmission line. The SQUIDs couple to the line via a mutual inductance only. (b) Further reduction of the system by including the SQUID influence into the relative, effective permeability $\mu_{\rm r,eff}$ seen by the line.}
 \label{fig:chain} 
\end{figure}

To retrieve the effective, relative permeability from $S_{21}$ for data presented in figures\,\ref{fig:ST}(b) and \ref{fig:mur2d}, we treat the system as a transmission line problem; the circuit diagram is shown in figure\,\ref{fig:chain}(a). $L'$ and $C'$ are the respective characteristic inductance and capacitance per unit length of transmission line. Then we simplify the unit cell by projecting the influence of the inductively coupled SQUIDs onto a change of the effective magnetic permeability seen by the transmission line and thus the line inductance (cf. Fig. \ref{fig:chain}(b)). 

\noindent
Using this model, it is a straight forward task to calculate the impedance matrix $Z$ and scattering matrix $S$ for the chain by cascading the unit cell's ABCD matrix.
\begin{eqnarray}
S & = & \left ( Z  - Z_0 \cdot \mathbbm{1}\right ) \left ( Z +  Z_0 \cdot \mathbbm{1}  \right )^{-1}\\
Z & = & \frac{1}{\tilde A_{\rm 21}} \left ( \begin{array}{cc}  \tilde A_{\rm 11}&  \det(\tilde A)\\ 1& \tilde A_{\rm 22} \end{array} \right )\\
\tilde A &=& A^{\rm N}
\end{eqnarray}
Here, $A = A\left ( \mur \right )$ and $\tilde A$ are the ABCD matrices of the unit cell and the chain of N unit cells, respectively. $Z_0$ is the port impedance on both sides.
Since the transmission line properties $L'$ and $C'$ of the CPW are known, the only free parameter is the effective, relative permeability $\mur$. The reverse process, however, (calculating $\mur$ from the scattering matrix) is more complicated because it involves choosing the correct root when solving a system of coupled nonlinear equations (see supplementary material). Similar to other well known methods \cite{nist}, this can be done by comparing the measured to the calculated phase delay. 
It should be noted that this retrieval method allows the reconstruction of $\mur$ using only one element of the scattering matrix which simplifies the experimental process significantly.

\section{Conclusion}
In this work, we have presented experimental results on a superconducting thin-film metamaterial that is tunable in situ by a weak dc magnetic field. As basic building blocks, we employed rf-SQUIDs embedded inside the gap of a coplanar waveguide. We have demonstrated a tunable, collective resonance of almost all 54 SQUIDs. Changing the resonance frequency hardly affects the quality of the resonance in wide frequency band. However, sample holder resonances distort the picture for some frequencies. Using only the transmission data $S_{21}$, we were able to extract the effective magnetic permeability from the measured data. These results show that we have created a material with tunable effective magnetic permeability in a frequency range between 10\,GHz and 14.5\,GHz. Further experiments will include an improved sample layout and environment in order to obtain results with higher quality factors and less parasitic resonances. 

\ack
The authors would like to acknowledge interesting and productive discussions with S. M. Anlage, I. Gabitov and G. Tsironis. This work was supported in part by the Ministry of Education and Science of the Russian Federation, and also by the Deutsche Forschungsgemeinschaft (DFG) and the State of Baden-W\"urttemberg through the DFG Center for Functional Nanostructures (CFN). Philipp Jung would like to acknowledge the financial support by the Helmholtz International Research School for Teratronics (HIRST), Susanne Butz would like to acknowledge the financial support by the Landesgraduiertenf\"orderung Baden-W\"urttemberg.

\end{document}


\title{Supplementary Material}
\section{Retrieval of $\mur$}
In order to retrieve the effective parameter $\mur$ mentioned in Section\,5, we describe the loaded waveguide as a lossless transmission line (see Fig.\,6). The ABCD matrix of one unit cell \cite{pozar} is
\begin{equation}
A = \left(\begin{array}{rr}
\frac{Z_{\rm L}}{Z_{\rm C}} + 1 & Z_{\rm L} \\
\frac{1}{Z_{\rm C}} & 1
\end{array}\right).
\end{equation}
Here, $Z_{\rm L} = i \omega \mur L' l$ is the impedance of the inductance and $Z_{\rm C} = 1/(i \omega C' l)$ is the impedance of the capacitance for a unit cell of length $l$. $L'$ and $C'$ are the inductance and capacitance per unit length of transmission line, respectively. The eigenvalues 
\begin{equation}
l_1 =  	
\frac{Z_{\rm L} + 2 \, Z_{\rm C} - \sqrt{Z_{\rm L}^{2} + 4 \, Z_{\rm L} Z_{\rm C}}}{2 \,
Z_{\rm C}}
, \qquad
l_2 = 
\frac{Z_{\rm L} + 2 \, Z_{\rm C} + \sqrt{Z_{\rm L}^{2} + 4 \, Z_{\rm L} Z_{\rm C}}}{2 \,
Z_{\rm C}}
\end{equation}
and eigenvectors
\begin{equation}
e_1 =\left(1,\,-\frac{l_{2} - 1}{Z_{\rm L}}\right)^T, \qquad e_2 = \left(1,\,-\frac{l_{1} - 1}{Z_{\rm L}}\right)^T
\end{equation}
of the ABCD matrix can be used to rewrite it in the form
\begin{eqnarray}
A = C \cdot \left ( \begin{array}{rr} l_1 & 0 \\ 0 & l_2 \end{array} \right )\cdot C^{-1} \\
{\rm with \quad} C =\left(\begin{array}{rr} e_1, e_2 \end{array}\right).
\end{eqnarray}
This simplifies the total ABCD matrix $\tilde A$ when cascading N unit cells:

\begin{equation}
\tilde A = A^N = C \cdot \left ( \begin{array}{rr} l_1^N & 0 \\ 0 & l_2^N \end{array} \right )\cdot C^{-1}.
\end{equation}
Using this equation and the well-known relation \cite{pozar} of $\tilde A$ to the scattering matrix $S$, we construct a system of four coupled, nonlinear equations

\begin{eqnarray}
0 & = & \left(\begin{array}{rr}
-\frac{{\left(l_{2} - 1\right)} l_{2}^{N} - {\left(l_{1} - 1\right)}
l_{1}^{N}}{l_{1} - l_{2}} & \frac{{\left(l_{1}^{N} -
l_{2}^{N}\right)} Z_{\rm L}}{l_{1} - l_{2}} \\
-\frac{{\left(l_{1} - 1\right)} l_{1}^{N} l_{2} - {\left(l_{1} -
1\right)} l_{1}^{N} - {\left({\left(l_{1} - 1\right)} l_{2} - l_{1} +
1\right)} l_{2}^{N}}{{\left(l_{1} - l_{2}\right)} Z_{\rm L}} &
\frac{{\left(l_{1} - 1\right)} l_{2}^{N} - l_{1}^{N} l_{2} +
l_{1}^{N}}{l_{1} - l_{2}}
\end{array}\right) - \\ 
\nonumber & &\left(\begin{array}{rr}
-\frac{{\left(S_{11} + 1\right)} S_{22} - S_{12} S_{21} - S_{11} - 1}{2
\, S_{21}} & \frac{{\left({\left(S_{11} + 1\right)} S_{22} - S_{12}
S_{21} + S_{11} + 1\right)} Z_{0}}{2 \, S_{21}} \\
\frac{{\left(S_{11} - 1\right)} S_{22} - S_{12} S_{21} - S_{11} + 1}{2
\, S_{21} Z_{0}} & -\frac{{\left(S_{11} - 1\right)} S_{22} - S_{12}
S_{21} + S_{11} - 1}{2 \, S_{21}}
\end{array}\right)
\end{eqnarray}
for the four variables $S_{11}$, $S_{12}$, $S_{22}$, and $\mur$ since $S_{21}$ is the measured quantity and all other parameters (like $L'$ and $C'$) are known from either design considerations or simulations.

\noindent
In general, this system of equations is not single-valued. There are several ways to sort out implausible solutions like requiring reciprocity, energy conservation or comparing the measured to the calculated phase delay.
\section{Data calibration \& reference planes}

In order to retrieve the effective, relative permeability from $S_{21}$ using the method described above, special care has to be taken to choose the correct calibration technique. In this approach, the reference planes of the calibration have to be just before and after the array of unit cells. In a real experiment, the best set of reference planes achievable for a full calibration of such a measurement is located at the microwave connectors closest to the sample at cryogenic temperatures. This, however, does not only require a more complex experimental setup \cite{Yeh2012}, it is also insufficient for our retrieval method. 

To solve this issue, we combine measurement and simulation to prepare the data for the retrieval algorithm. First of all, we divide the experimental setup into three parts, each of which can be described individually by a two-port scattering matrix:\\

\begin{tabular}{rp{.85\textwidth}}
$S^{\rm in}$& describes the input part of the setup, from port 1 of the network analyzer to the beginning of the SQUID loaded section of the CPW.\\
$S^{\rm stl}$& describes the SQUID loaded section of the CPW.\\
$S^{\rm out}$& describes the output part of the setup from the end of the SQUID loaded section of the CPW to port 2 of the network analyzer.\\
\end{tabular}\\

\noindent
We measure the total transmission from VNA port 1 to port 2 which can be written\footnote{By multiplying the transfer matrices corresponding to these scattering matrices. \cite{pozar}} as
\begin{equation}
S^{\rm tot}_{21} = 
-\frac{S^{\rm in}_{21} S^{\rm stl}_{21}
S^{\rm out}_{21}}{{\left(S^{\rm in}_{22} S^{\rm stl}_{12} S^{\rm stl}_{21} -
{\left(S^{\rm in}_{22} S^{\rm stl}_{11} - 1\right)} S^{\rm stl}_{22}\right)}
S^{\rm out}_{11} + S^{\rm in}_{22} S^{\rm stl}_{11} - 1}
.
\end{equation}
If we assume that the waveguides leading to the sample are well matched (i.e. no reflections at the waveguide connections to the sample: $S^{\rm in}_{22} = S^{\rm out}_{11} = 0$), the expression simplifies to
\begin{equation}
S^{\rm tot}_{21} = S^{\rm in}_{21} S^{\rm stl}_{21}S^{\rm out}_{21}
.
\end{equation}
Clearly, all of the above matrix elements are frequency dependent. $S^{\rm stl}_{21}$, however, also depends on the dc magnetic field. It can be decomposed into two factors by assuming $ S^{\rm stl}_{21} = \tilde S^{\rm stl}_{21} \left ( \omega \right ) \cdot \alpha \left ( \omega, \Phi_{\rm e0} \right )$.\\

\begin{tabular}{rp{.8\textwidth}}
$\tilde S^{\rm stl}_{21} \left ( \omega \right ) $& is the transmission through the loaded waveguide section without the effect of the SQUID resonance. Consequently, this factor does not depend on magnetic field.\\
$\alpha \left ( \omega, \Phi_{\rm e0} \right )$& describes the change in the transmission through the loaded waveguide due to the SQUID resonance. Therefore, this factor is frequency and field dependent.\\
\end{tabular}\\

\noindent
The total transmission then reads
\begin{equation}
S^{\rm tot}_{21}\left ( \omega, \Phi_{\rm e0}\right ) = S^{\rm in}_{21}\left ( \omega\right ) \tilde S^{\rm stl}_{21}\left ( \omega\right ) \alpha \left ( \omega, \Phi_{\rm e0}\right )  S^{\rm out}_{21}\left ( \omega\right ) .
\end{equation}

\noindent
The goal of the calibration is to extract the term $S^{\rm stl}_{21}$ from the measured $S^{\rm tot}_{21}$ since the former is the quantity needed for the effective $\mur$ retrieval. 
If we restrict our investigation to a limited frequency range bounded by $\omega_{\rm min}$ and $\omega_{\rm max}$, we can usually find a value of the flux $\Phi_{\rm e0} = \Phi_{\rm cal}$ for which $\alpha \left ( \omega, \Phi_{\rm cal}\right ) \approx 1 $.
Thus, by dividing all the measured data in the specified frequency range by the corresponding value at the calibration flux\footnote{Choosing this ``calibration" flux effectively limits the validity of the measured data to the mentioned frequency range.}, we can extract $\alpha$:
\begin{equation}
S^{\rm tot,cal}_{21}\left ( \omega, \Phi_{\rm e0}\right ) =\frac{ S^{\rm tot}_{21}\left ( \omega, \Phi_{\rm e0}\right )}{ S^{\rm tot}_{21}\left ( \omega, \Phi_{\rm cal}\right )} \approx \alpha \left ( \omega, \Phi_{\rm e0}\right )
\end{equation}

\noindent
As a last step, we have to reconstruct $S^{\rm stl}_{21}$ from the calibrated data $S^{\rm tot,cal}_{21}$. Therefore, we simulate the transmission through the loaded waveguide without the SQUID resonance $\tilde S^{\rm stl}_{21}\left ( \omega\right )$ and multiply it by the calibrated data:
\begin{equation}
S^{\rm stl,reconstructed}_{21}\left (\omega, \Phi_{\rm e0} \right ) = S^{\rm tot,cal}_{21}\left ( \omega, \Phi_{\rm e0}\right ) \cdot \tilde S^{\rm stl, simulated}_{21}\left ( \omega\right )
\end{equation}
This result can then be used in the algorithm outlined in the last section. It should be stressed, however, that this method is only valid if the effect of the SQUID resonance at the calibration flux is negligible in the frequency range of interest.